\documentstyle[prl,twocolumn,aps]{revtex}

\def\be{\begin{equation}}
\def\ee{\end{equation}}
\def\bea{\begin{eqnarray}}
\def\nn{\nonumber}
\def\eea{\end{eqnarray}}

\begin{document}

\draft

\wideabs{
\title{On the semiclassical Einstein-Langevin equation}
\author{Rosario Mart\'{\i}n and Enric Verdaguer}
\address{Departament de F\'{\i}sica Fonamental, Universitat de
        Barcelona, Av.~Diagonal 647, \mbox{08028 Barcelona}, Spain}
\date{\today}
\maketitle

\begin{abstract}
We introduce a semiclassical Einstein-Langevin equation as a
consistent dynamical equation 
for a first order perturbative correction to semiclassical gravity.
This equation includes 
the lowest order quantum stress-energy fluctuations 
of matter fields as a source of classical stochastic
fluctuations of the gravitational field.
The Einstein-Langevin equation is explicitly solved around
one of the simplest solutions of semiclassical gravity:
Minkowski spacetime with a conformal scalar field in its vacuum
state. We compute the 
two-point correlation function of the linearized Einstein tensor.
This calculation illustrates the possibility of obtaining some
``non-perturbative'' behavior for the induced gravitational
fluctuations that cannot be obtained in perturbative quantum gravity.
\end{abstract}


}

\narrowtext


Semiclassical gravity describes the interaction of the
gravitational field, which is treated classically, with quantum matter
fields. The theory
is mathematically well defined and
fairly well understood, at least for linear matter 
fields \cite{wald,wald77}.
The equation of motion for the classical metric is the 
semiclassical Einstein equation, which gives the back reaction of the
matter fields on the spacetime;
it is a generalization of
the Einstein equation where the source  
is the expectation value in some quantum state
of the matter stress-energy tensor operator.

In the absence of a complete quantum theory of gravity interacting
with matter fields
from which the semiclassical theory can be derived,
the scope and limits of semiclassical gravity are less well
understood. It seems clear, however, that it should not be valid
unless gravitational fluctuations are negligibly small
\cite{flanagan,ashtekar}. 
This condition may break down when the matter stress-energy has
appreciable quantum fluctuations \cite{wald,wald77},
given that a quantum metric operator should couple to
the stress-energy operator of matter and, thus, 
fluctuations in the stress-energy of matter would induce
gravitational fluctuations \cite{ford82}.
In recent years, a number of examples have been studied,
including some quantum fields in cosmological models and even in flat
spacetimes with non-trivial topology, where, for some states 
of the fields, the stress-energy tensor have significant
fluctuations \cite{stress-en_fluctu}.
It thus seems of interest to try to generalize the semiclassical
theory to account for such fluctuations.

In this paper we propose a generalization of semiclassical gravity
based on a semiclassical Einstein-Langevin equation, which describes
the back reaction on the spacetime metric of the lowest order 
stress-energy fluctuations. This results in an effective theory which
predicts linear stochastic corrections to the semiclassical
metric and may be applicable when gravitational fluctuations of
genuine quantum nature can be neglected. 
It should be stressed that the gravitational fluctuations predicted by
this stochastic semiclassical theory are ``passive'' rather than
``active'', {\it i.e.}, they are induced by the matter field
stress-energy fluctuations \cite{ford95}. 
Thus, we cannot expect that stochastic
semiclassical gravity gives a satisfactory description of
gravitational fluctuations in all situations. 
In spite of that, this theory may have
a number of interesting applications in the physics of the early
universe and of black holes, where one may expect significant
matter stress-energy fluctuations.

The idea which motivates such stochastic theory is that, 
in the transition from the full quantum
regime to the semiclassical one, there should be some mechanism which
effectively suppresses
the quantum interference effects in the gravitational field.  
Once such decoherence process has taken place, gravity might undergo
an intermediate regime in which it would
fluctuate, but it would do so classically. 
If such a regime exists, an 
effective probabilistic description of the
gravitational field by a stochastic metric field may be possible
\cite{eff_equations}.

Another motivation for our work is to connect with some recent results
on equations of the Langevin-type
that have appeared in the context of
semiclassical cosmology and which predict stochastic perturbations
around some cosmological backgrounds \cite{Langevin,cv}.
Although physically motivated \cite{hu}, 
the derivation of these equations,
obtained by functional methods, is formal and doubts may be raised on
the physical significance of the predicted fluctuations. 
Our approach is complementary to these functional methods, 
it allows to write these equations in a general form
and it links the source of such metric fluctuations with the
matter stress-energy fluctuations.

The idea, when implemented in the framework of a
perturbative approach around semiclassical gravity, is quite simple.
We start realizing that, for 
a given solution of semiclassical gravity, one can
associate the lowest order matter stress-energy fluctuations to a
classical stochastic tensor field. Then, we seek a
consistent equation which incorporates this stochastic tensor as
a source of a first order correction to semiclassical gravity. 
It is important to remark that, 
even if this approach differs from
those based on functional methods, the semiclassical 
Einstein-Langevin equation introduced here can actually be
formally derived using those methods. The details of such derivation
will be given in a subsequent paper \cite{mv}.

As a simple application of stochastic semiclassical gravity,
we have explicitly solved the Einstein-Langevin equation
around a solution of semiclassical gravity
consisting on Minkowski spacetime with a conformal scalar field in the
Minkowskian vacuum state. We have computed the
two-point correlation function for the induced 
linearized Einstein tensor. 
Even if, as expected, such correlation function has negligible values
for points separated by scales larger than the Planck scales, 
it is interesting to compare our result with the one obtained from a
perturbative approach to quantum gravity \cite{donoghue}.
The result hints at the possibility of obtaining some
non-perturbative (in the Planck length) 
behavior for the induced gravitational fluctuations
which could not be obtained in perturbative quantum gravity.
Throughout we use the $(+++)$ sign conventions and work in units in
which $c=\hbar =1$.


\paragraph*{The semiclassical Einstein-Langevin equation.---}
We start with the semiclassical Einstein equation. Let
$(M,g_{ab})$ be a globally hyperbolic four-dimensional
spacetime and consider a linear quantum field $\Phi$ on 
this background. Working in the 
Heisenberg picture, let $\hat{\Phi}[g]$ 
and $\hat{\rho}[g]$ be respectively
the field operator and the density operator describing the state of
the field. Such state will be assumed to be physically acceptable
in the sense of Ref.~\cite{wald}.

The set $(M,g_{ab},\hat{\Phi},\hat{\rho})$ is a solution of 
semiclassical gravity if it satisfies
the semiclassical Einstein equation:
\be
{1\over 8 \pi G} \left( G_{ab}[g]+ \Lambda g_{ab} \right)-
2  \left( \alpha A_{ab}+\beta B_{ab} \right)\hspace{-0.3ex}[g]=
\left\langle\hat{T}^{R}_{ab}\right\rangle \![g], 
\label{semiclassical Einstein eq}
\ee
where $\hat{T}^{R}_{ab}[g]$ is the renormalized
stress-energy tensor operator for the field $\hat{\Phi}[g]$,
which satisfies the corresponding field equation on 
the spacetime $(M,g_{ab})$, 
and the expectation value is taken in the state described by 
$\hat{\rho}[g]$ \cite{wald,wald77}. 
In the above equation, $1/G$, $\Lambda /G$, $\alpha$ and $\beta$ are
renormalized coupling constants,
$G_{ab}$ is the Einstein tensor, and $A_{ab}$ and
$B_{ab}$ are the local curvature tensors obtained by 
functional derivation with
respect to the metric of the action terms corresponding 
to the Lagrangian
densities $C_{abcd}C^{abcd}$ and $R^2$, respectively, 
where $C_{abcd}$ is the Weyl tensor and $R$ is the scalar curvature.
A classical stress-energy tensor can also be added to the right hand
side of Eq.~(\ref{semiclassical Einstein eq}), but, for simplicity, 
we shall ignore this term.

Given a solution of semiclassical gravity, the matter stress-energy
tensor will in general have quantum fluctuations. To lowest order,
such fluctuations may be described by the following bi-tensor, which
we call noise kernel, 
\be
8 N_{abcd}(x,y) \equiv \bigl\langle  \bigl\{
 \hat{t}_{ab}(x) , \,
 \hat{t}_{cd}(y)
 \bigr\} \bigr\rangle [g],
\label{noise}
\ee
where $\{ \; , \: \}$ means the anticommutator
and $\hat{t}_{ab} \equiv \hat{T}_{ab}-
 \langle \hat{T}_{ab} \rangle$, 
where $\hat{T}_{ab}[g]$ denotes the unrenormalized stress-energy
``operator'' 
[we use the word ``operator'' for $\hat{T}_{ab}$ in a formal sense;
it should be understood that the matrix elements of this ``operator''
are suitably regularized and that the regularization is removed
after computing the right hand side of (\ref{noise})].
For a linear matter field, this noise kernel is
free of ultraviolet divergencies,
and the ``operator'' $\hat{t}_{ab}$ in (\ref{noise})
can be replaced by the
operator $\hat{T}^{R}_{ab}-\langle \hat{T}^{R}_{ab} \rangle$.

We want now to introduce an equation in which the 
stress-energy fluctuations described by (\ref{noise}) 
are the source of classical gravitational fluctuations,
as a perturbative correction to semiclassical gravity. 
Thus, we assume that the gravitational field is 
described by  $g_{ab}+h_{ab}$,
where $h_{ab}$ is a linear perturbation 
to the metric $g_{ab}$, solution of
Eq.~(\ref{semiclassical Einstein eq}). 
The renormalized stress-energy
operator and the density operator
describing the state of the field
will be denoted as $\hat{T}^{R}_{ab}[g+h]$ and
$\hat{\rho}[g+h]$, respectively, and 
$\left\langle\hat{T}^{R}_{ab}\right\rangle \![g+h]$
will be the corresponding expectation value.

In order to write 
an equation which describes the dynamics
of the metric perturbation $h_{ab}$, 
let us introduce a Gaussian stochastic tensor field $\xi_{ab}$
characterized by the following correlators:
\be
\left\langle\xi_{ab}(x) \right\rangle_c = 0,  
\hspace{6ex}
\left\langle\xi_{ab}(x)\xi_{cd}(y) \right\rangle_c = N_{abcd}(x,y),
\label{correlators}
\ee
where 
$\langle \hspace{1.5ex} \rangle_c$ means statistical 
average. 
Note that the two-point correlation function of a stochastic tensor
field $\xi_{ab}$ must be a symmetric positive semi-definite
real bi-tensor field (since, obviously,  
$\left\langle\xi_{ab}(x)\xi_{cd}(y) \right\rangle_c =
\left\langle\xi_{cd}(y)\xi_{ab}(x) \right\rangle_c$). 
Since $\hat{T}^{R}_{ab}$ is self-adjoint, 
it is easy to see from the definition
(\ref{noise}) that $N_{abcd}(x,y)$ satisfies all these conditions.
Therefore, relations (\ref{correlators}), with the cumulants of higher
order being zero, do truly characterize a stochastic tensor field 
$\xi_{ab}$.
One could also seek higher order corrections which would take into
account higher order stress-energy fluctuations, but
we stick, for simplicity, to the lowest order. 
The simplest equation which can incorporate in a consistent way the
stress-energy fluctuations described by $N_{abcd}(x,y)$
as the source of metric fluctuations is
\bea
&&{1\over 8 \pi G} \Bigl( G_{ab}[g+h]+ 
\Lambda\left(g_{ab}+h_{ab}\right) \Bigr)    \nn \\
&& \hspace{3.5ex}
-\,2 \left( \alpha A_{ab}+\beta B_{ab} \right)\hspace{-0.3ex}
[g+h]=\left\langle
\hat{T}^{R}_{ab}\right\rangle \![g+h] +2 \xi_{ab}, 
\label{Einstein-Langevin eq}
\eea
which must be understood to linear order in
$h_{ab}$. 
This is the semiclassical Einstein-Langevin equation, which 
gives a first order correction to semiclassical gravity.
Notice that, in writing Eq.~(\ref{Einstein-Langevin eq}), we are
implicitly assuming that $h_{ab}$ is also a stochastic tensor field.

We must now ensure that this equation is consistent and, thus, it can
truly describe the dynamics of metric perturbations.
Note that the term $\xi_{ab}$ in Eq.~(\ref{Einstein-Langevin eq}) 
is not dynamical, {\it i.e.}, it does not depend on $h_{ab}$,
since it is defined through
the semiclassical metric $g_{ab}$ by the correlators
(\ref{correlators}). Being the source of the metric perturbation
$h_{ab}$, this 
term is of first order in perturbation theory around semiclassical
gravity. Let us now see that $\xi_{ab}$
is covariantly conserved up to first order in perturbation
theory, in the sense that the stochastic vector field 
$\bigtriangledown^{a} \xi_{ab}$ is deterministic 
and represents with certainty the zero vector field 
on $M$ ($\bigtriangledown^{a}$ means the covariant derivative 
associated to the metric $g_{ab}$).
Using that 
$\bigtriangledown^{a} \hat{T}^{R}_{ab}[g]\!=\!0$, we have
$\bigtriangledown^{a}_{\! \mbox{}_{x}}\!
N_{abcd}(x,y)\!=\!0$ and, from the covariant derivative of the
correlators (\ref{correlators}), we get 
$\langle \bigtriangledown^{a}
\xi_{ab} \rangle_c\!=\!0$ and 
$\langle \bigtriangledown^{a}_{\! \mbox{}_{x}}
\xi_{ab}(x) \!\!
\bigtriangledown^{c}_{\! \mbox{}_{y}} \!\!
\xi_{cd}(y) \rangle_c\!=\!0$.
It is thus consistent to include the
term $\xi_{ab}$ in the right hand side of 
Eq.~(\ref{Einstein-Langevin eq}). 
Note that for a
conformal matter field, {\it i.e.}, a field whose classical action is
conformally invariant, 
the stochastic source $\xi_{ab}$
is ``traceless'' up to first order in perturbation theory.
That is, $g^{ab}\xi_{ab}$ is
deterministic and  represents with certainty the zero scalar
field on $M$. In fact, from the
trace anomaly result, which states that $g^{ab} \hat{T}^{R}_{ab}[g]$ 
is in this case a local c-number functional of $g_{cd}$ times the
identity operator, we have that
$g^{ab}(x) N_{abcd}(x,y)=0$. It then follows from (\ref{correlators})
that $\langle g^{ab}\xi_{ab} \rangle_c=0$ and 
$\langle  g^{ab}(x)\xi_{ab}(x)g^{cd}(y)\xi_{cd}(y) \rangle_c=0$.
Hence, in the case of a conformal matter field, 
the stochastic source gives no correction to the trace anomaly.

Since Eq.~(\ref{Einstein-Langevin eq}) gives a linear stochastic
equation for $h_{ab}$ with an inhomogeneous term $\xi_{ab}$,
a solution can be formally written as a functional
$h_{ab}[\xi]$ of the stochastic source $\xi_{cd}$. 
Such a solution can be characterized by 
the whole family of its correlation functions. 
By taking the average of
Eq.~(\ref{Einstein-Langevin eq}), one sees that the 
metric $g_{ab}+\left\langle h_{ab} \right\rangle_c$,
must be a solution of the semiclassical Einstein equation linearized
around $g_{ab}$.    
For the solutions of Eq.~(\ref{Einstein-Langevin eq}) we have the
gauge freedom 
$h_{ab} \rightarrow 
h'_{ab}\equiv h_{ab}+ \bigtriangledown_{\!a}
\zeta_{b}+\bigtriangledown_{\!b} \zeta_{a}$, 
where $\zeta^{a}$ is any stochastic vector field on $M$ which is a
functional of the Gaussian stochastic field
$\xi_{cd}$, and $\zeta_{a} \equiv g_{ab}\zeta^{b}$,
so that $h_{ab}$ and $h'_{ab}$ are physically equivalent
solutions.


\paragraph*{Vacuum fluctuations in flat spacetime.---}
As an example, we shall now consider the class of 
trivial solutions of
semiclassical gravity. Each of such solutions consists of 
Minkowski spacetime,
$({\rm I\hspace{-0.4 ex}R}^{4},\eta_{ab})$, a linear matter field, 
and the usual Minkowskian vacuum state for this field,
$\hat{\rho}[\eta]=|0 \rangle \langle 0|$. As it is well known, 
we can always choose a renormalization scheme in which 
$\langle 0|\, \hat{T}^{R}_{ab}\, |0 \rangle [\eta]=0$. 
Thus, each of the above sets is a solution to 
Eq.~(\ref{semiclassical Einstein eq}) with $\Lambda\!=\!0$.
Note that, although the
vacuum $|0 \rangle$ is an eigenstate of the total four-momentum
operator which can be defined in Minkowski spacetime, such state
is not an eigenstate of $\hat{T}^{R}_{ab}[\eta]$. Hence, even in 
these trivial solutions, quantum fluctuations of the matter
stress-energy are present, and 
the noise kernel defined in (\ref{noise}) does not vanish.
This fact leads to consider the Einstein-Langevin
correction to the trivial solutions of semiclassical gravity
\cite{mv}.
For a massless field, these vacuum stress-energy fluctuations 
should be characterized by the only scale available,
the Planck length, and, thus, they should be very small on macroscopic
scales, unlike the cases considered in 
Refs.~\cite{stress-en_fluctu}.

The corresponding Einstein-Langevin equation becomes simpler
when the matter field is a
massless conformally coupled real scalar field. 
In the global
inertial coordinate system $\{x^\mu \}$,  
the components of the flat metric are
simply $\eta_{\mu\nu}={\rm diag}(-1,1,1,1)$. 
We shall use $G^{\scriptscriptstyle (1)}_{\mu\nu}$,
$A^{\scriptscriptstyle (1)}_{\mu\nu}$ and
$B^{\scriptscriptstyle (1)}_{\mu\nu}$ to denote, respectively,
the components of the tensors $G_{ab}$, $A_{ab}$ and $B_{ab}$
linearized around the flat metric. These tensor components can be
written in terms of $G^{\scriptscriptstyle (1)}_{\mu\nu}$ as
\[
A^{{\scriptscriptstyle (1)}}_{\mu\nu}=
{2 \over 3} \, ({\cal F}_{\mu\nu} 
G^{{\scriptscriptstyle (1)}}\mbox{}^{\alpha}_\alpha
-{\cal F}^{\alpha}_\alpha 
G^{{\scriptscriptstyle (1)}}_{\mu\nu}),
\hspace{6ex}
B^{{\scriptscriptstyle (1)}}_{\mu\nu}=
2 \hspace{0.2ex} {\cal F}_{\mu\nu} 
G^{{\scriptscriptstyle (1)}}\mbox{}^{\alpha}_\alpha,
\]
where ${\cal F}_{\mu\nu}$ is the differential operator 
${\cal F}_{\mu\nu} \equiv \eta_{\mu\nu} \Box
-\partial_\mu \partial_\nu$.
It is also convenient to introduce the Fourier transform,
$\tilde{f}(p)$, of a field
$f(x)$ as 
$f(x) \equiv (2\pi)^{-4} \!
\int \! d^4 p  \,
e^{i px}\, \tilde{f}(p)$.
Eq.~(\ref{Einstein-Langevin eq}) reduces in this case to 
\bea
&&{1\over 8 \pi G} \,
G^{\scriptscriptstyle (1)}_{\mu\nu}(x)
- 2 \left( \bar{\alpha} 
A^{\scriptscriptstyle (1)}_{\mu\nu}
+\bar{\beta}
B^{\scriptscriptstyle (1)}_{\mu\nu} \right)\!(x)   \nn \\
&& \hspace{3.5ex}
+\int\! d^4y \, H(x\!-\!y;\mu^2) \,
A^{\scriptscriptstyle (1)}_{\mu\nu}(y)
=2 \xi_{\mu\nu}(x),
\label{Einstein-Langevin eq 2}
\eea
where $\bar{\alpha} \equiv \alpha+1/(3600 \pi^2)$,
$\bar{\beta} \equiv \beta-1/(34560 \pi^2)$, and
$\tilde{H}(p;\mu^2) \equiv 
\left( \ln \left| \hspace{0.2ex} p^2/  \mu^2 \hspace{0.2ex}\right|
- i \pi \, {\rm sign}\,p^0 \; \theta (-p^2) \right)/ (1920 \pi^2)$,  
being $\mu$ a renormalization mass scale \cite{cv}. 
The components
$\xi_{\mu\nu}$ of the stochastic source tensor satisfy in this
case 
\[
\langle\xi_{\mu\nu}(x)\xi_{\alpha\beta}(y) \rangle_c
= {1 \over 6} \, {\cal F}_{\mu\nu\alpha\beta}^{x}\,
 N(x-y),
\]
where 
${\cal F}_{\mu\nu\alpha\beta} \equiv
3 \hspace{0.2ex}
{\cal F}_{\mu (\alpha}{\cal F}_{\beta )\nu}-
{\cal F}_{\mu\nu}{\cal F}_{\alpha\beta}$ and
$\tilde{N}(p) \equiv \theta (-p^2)/(1920 \pi)$.

Equations (\ref{Einstein-Langevin eq 2}) can be solved for
the components $G^{\scriptscriptstyle (1)}_{\mu\nu}$ of the linearized
Einstein tensor. 
Using a procedure similar to that
described in the Appendix of Ref.~\cite{flanagan}, 
we find the family of solutions which can be written as a
linear functional of the stochastic source 
and whose Fourier transform 
$\tilde{G}^{\scriptscriptstyle (1)}_{\mu\nu}(p)$ depends locally on
$\tilde{\xi}_{\alpha\beta}(p)$. 
Each of such solutions is a Gaussian stochastic field
and, thus, it can be completely characterized by its one-point and
two-point correlation functions. The one-point correlation functions,
{\it i.e.}, the averages 
$\langle G^{\scriptscriptstyle (1)}_{\mu\nu}\rangle_c$, 
are solution of the linearized semiclassical Einstein equations
obtained by averaging Eqs.~(\ref{Einstein-Langevin eq 2});
solutions to these equations were first found by Horowitz 
\cite{horowitz}.
We can then compute the two-point correlation functions
${\cal G}_{\mu\nu\alpha\beta}(x,x^{\prime}) \equiv 
\langle G^{ \scriptscriptstyle (1)}_{\mu\nu}(x)
G^{\scriptscriptstyle (1)}_{\alpha\beta}(x^{\prime}) \rangle_c
-\langle G^{\scriptscriptstyle (1)}_{\mu\nu}(x)\rangle_c 
\langle G^{\scriptscriptstyle (1)}_{\alpha\beta}(x^{\prime}) 
\rangle_c$ \cite{mv}.
These correlation functions are invariant under gauge
transformations of the metric perturbations and give a measure of the
induced gravitational fluctuations in the present context, 
to the extent that these 
fluctuations can be described by exact solutions to 
Eqs.~(\ref{Einstein-Langevin eq 2}) (the result would be different if
one applied some ``reduction of order'' procedure \cite{flanagan}). 
We get
\be
{\cal G}_{\mu\nu\alpha\beta}(x,x^{\prime}) =
{\pi \over 45} \, G^2 {\cal F}_{\mu\nu\alpha\beta}^{x} \,
{\cal G} (x-x^{\prime}),
\label{correlation functions}
\ee
where $\tilde{{\cal G}}(p)= \theta(-p^2) \left|\hspace{0.2ex} 
1+16 \pi G\hspace{0.2ex} p^2\hspace{0.2ex} \tilde{H}(p;\bar{\mu}^2) 
\hspace{0.2ex} \right|^{-2}$, with 
$\bar{\mu} \equiv \mu\, \exp (1920 \pi^2 \bar{\alpha})$.
Introducing the function 
$\varphi (\chi ; \lambda ) \equiv \left[ 1-\chi 
\ln \hspace{-0.2ex}\left(\lambda \chi /e\right)\right]^2
+\pi^2 \chi^2$, 
with $\chi \geq 0$ and $\lambda > 0$,
${\cal G}(x)$ can be written as
\bea
&&{\cal G}(x) = {(120 \pi)^{3/2} \over 2 \pi^3 L_P^3}  \,
{1 \over |{\bf x}|} \!\int_{0}^{\infty}\!\! d|{\bf q}| \hspace{0.2ex}
|{\bf q}|
\sin \!\left[ { \sqrt{120 \pi} \over L_P}\,
|{\bf x}| \hspace{0.2ex} |{\bf q}| \right]  
\nn   \\
&& \hspace{10ex} \times \int_{0}^{\infty}\! dq^0 
\cos \!\left[ { \sqrt{120 \pi} \over L_P} \,
x^0 q^0 \right] {\theta(-q^2) \over \varphi (-q^2; \lambda )}, 
\label{G}
\eea
where $x^{\mu}=(x^0,{\bf x})$ and $q^{\mu}=(q^0,{\bf q})$,
$\lambda \equiv 120 \pi e /(L_P^2 \bar{\mu}^2)$, and 
$L_P \equiv \sqrt{G}$ is the Planck length. If we assume that
$\bar{\mu} \leq L_P^{-1}$, then $\lambda > 10^3$.
Let us consider the case when 
$(x-x^{\prime})^\mu$ is spacelike, then we can choose
$(x-x^{\prime})^\mu=(0,{\bf x}-{\bf x}^\prime)$ and
${\cal G}_{\mu\nu\alpha\beta}(x,x^{\prime})$
will be a function of ${\bf x}-{\bf x}^\prime$ only.
From (\ref{correlation functions}) and (\ref{G}), one can see that
${\cal G}_{000i}({\bf x}-{\bf x}^\prime)=
{\cal G}_{0ijk}({\bf x}-{\bf x}^\prime)=0$. The remaining components 
are non-null and can be explicitly computed performing the
integrals with some approximations. The result depends on the constant
$\kappa \equiv \ln \hspace{-0.2ex}
\left( \lambda \chi_0(\lambda)/e \right)$, being $\chi_0(\lambda)$ 
the value of $\chi$ where
$\varphi (\chi ; \lambda )$ has its minimum (this can be
found by solving the equation 
$\pi^2 \chi_0 = \left[1-\chi_0 \ln( \lambda \chi_0 /e)\right] 
\left[1+ \ln( \lambda \chi_0 /e) \right]$
numerically).  For ${\bf x} \neq {\bf x}^\prime$, we get 
\[
{\cal G}_{0000}({\bf x}-{\bf x}^\prime) \simeq 
{2 \over 3 \pi} \, {a \hspace{0.2ex} b^6 \over L_P^4} \, 
{1 \over \sigma^2} \, 
e^{-\sigma} \!
\left[1+{4 \over \sigma}+ {12 \over \sigma^2}+
{24 \over \sigma^3}+{24 \over \sigma^4} \right]\hspace{-0.3ex},
\]
where 
$\sigma \equiv b \, |{\bf x}-{\bf x}^{\prime}|/L_P$,
$a \equiv 1+ (2/ \pi) 
\arctan (\kappa / \pi)$
and 
$b \equiv  (4 a/ \pi^2) \!
\left[15 \pi \!
\left(\sqrt{\kappa^2 +\pi^2}- \kappa \right) \right]^{1/2}$.
We should remark that this is not an expansion in $\sigma$. Similar
results are  obtained for the other non-null components. 
From these results, one can conclude that, as expected in this case, 
there are negligibly small correlations for the Einstein tensor at
points separated by distances large compared to the Planck
length. Thus, at such scales, the semiclassical approach is
satisfactory enough to describe the dynamics of gravitational
perturbations in Minkowski spacetime \cite{horowitz,flanagan}.
Deviations from semiclassical gravity start to be important at
Planckian scales. At such scales, however, 
gravitational fluctuations of genuine quantum nature cannot be
neglected and, thus, the classical description based on the
Einstein-Langevin equation would break down.
Note, however, the factor $e^{-\sigma}$ in our result, which is
non-analytic in the Planck length and gives a characteristic
correlation length of the order of $L_P$. This kind of behavior 
cannot be obtained from 
a perturbative approach to quantum gravity, 
in which one expands physical quantities as a
power series in $L_P^2$ \cite{donoghue}. 
Actually, if we had naively assumed that 
(\ref{correlation functions}) can be Taylor expanded in
$L_P^2$ (or if we had applied the reduction of order procedure), 
to leading order, and for 
$x \neq  x^\prime$, we would have obtained
${\cal G}_{\mu\nu\alpha\beta}(x,x^{\prime}) \sim
L_P^4 {\cal F}_{\mu\nu\alpha\beta}^{x} \, (x-x^{\prime})^{-4}$, 
a result qualitatively similar to that
of the analogous two-point function at one loop
in perturbative quantum gravity.
For solutions of semiclassical gravity with other scales present
apart from the Planck scales, induced gravitational fluctuations may
occur on macroscopic correlation scales. 
In such cases, the above results suggest that 
stochastic semiclassical gravity might yield 
physically relevant results which cannot be obtained from a
calculation in the framework of perturbative quantum gravity.


We are grateful to Esteban Calzetta, Jaume Garriga,
Bei-Lok Hu, Ted Jacobson and Albert Roura
for very helpful suggestions and discussions. 
This work has been partially supported by the 
CICYT Research Project number
\mbox{AEN95-0590}, and the European Project number
\mbox{CI1-CT94-0004}.


\end{document}